%% file: main.tex
\documentclass[10pt,twocolumn]{article}

\usepackage[a4paper,top=1.8cm,bottom=1.9cm,left=1.7cm,right=1.7cm,columnsep=0.7cm]{geometry}
\usepackage[T1]{fontenc}
\usepackage[utf8]{inputenc}
\usepackage{lmodern}
\usepackage{microtype}
\usepackage{graphicx}
\usepackage{booktabs}
\usepackage{tabularx}
\usepackage{threeparttable}
\usepackage{array}
\usepackage{multirow}
\usepackage{amsmath,amssymb,mathtools}
\usepackage{siunitx}
\usepackage{xcolor}
\usepackage{enumitem}
\usepackage{caption}
\usepackage{subcaption}
\usepackage{float}
\usepackage{placeins}
\usepackage{titlesec}
\usepackage{fancyhdr}
\usepackage{csquotes}
\usepackage[
  backend=biber,
  style=numeric-comp,
  sorting=none,
  sortcites=true,
  maxbibnames=99,
  doi=true,
  url=false,
  isbn=false,
  giveninits=true
]{biblatex}
\usepackage[colorlinks=true,allcolors=blue!55!black]{hyperref}
\hypersetup{
  pdftitle={The Generative AI Gold Rush in Theoretical and Computational Research},
  pdfauthor={Xiaoshn Nee, Haobo Zhong, and Xiaomin Ni},
  pdfkeywords={generative AI, mathematics, theoretical physics, computational research, scientific publishing, synthetic control, causal inference}
}

\input{generated_descriptive_metrics.tex}

\input{generated_causal_metrics.tex}
\input{generated_author_metrics.tex}
\input{generated_subfield_metrics.tex}

\newcommand{\arxiv}{arXiv}
\newcommand{\GenAI}{generative AI}
\newcommand{\LLM}{large language model}
\newcommand{\SCM}{synthetic control method}

\newcommand{\ind}{\mathbb{1}}
\newcommand{\code}[1]{\texttt{#1}}
\newcolumntype{Y}{>{\raggedright\arraybackslash}X}

\titleformat{\section}[hang]{\normalfont\Large\bfseries\filright}{\thesection}{0.8em}{}
\titleformat{\subsection}[hang]{\normalfont\large\bfseries\filright}{\thesubsection}{0.8em}{}
\setlist{nosep,leftmargin=*}
\newcommand{\runningtitle}{The Generative AI Gold Rush in Research}
\fancypagestyle{plain}{%
  \fancyhf{}
  \fancyfoot[C]{\thepage}
  }

\title{\textbf{The Generative AI Gold Rush in Theoretical and Computational Research}}
\author{%
Xiaoshn Nee$^{a}$\thanks{Corresponding author. Email: \href{mailto:nixsh3@gmail.com}{nixsh3@gmail.com}}, Haobo Zhong$^{b}$, Xiaomin Ni$^{c}$\\[0.65em]
{\small $^{a}$Independent researchers}\\
{\small $^{b}$HSBC Business School, Peking University, Shenzhen City, Guangdong 518055, China}\\
{\small $^{c}$Artificial Intelligence Research Institute, Shenzhen University of Advanced Technology, Shenzhen, China}}
\date{3 September 2026}

\begin{document}
\raggedbottom
\maketitle

\begin{abstract}
Generative AI is changing the production conditions of theoretical and computational research, but its system level effects require measures that separate platform growth, field specific divergence, and production structure. We assemble 2,080 monthly observations for twenty \arxiv{} archives from January 2018 through August 2026 and a separate pseudonymized Mathematics author panel. A regularized convex synthetic control fitted through December 2025 identifies the January--August 2026 anomaly, while spatial placebos, prior year pseudo holdouts, donor refits, and alternative preperiods assess comparative robustness. Mathematics recorded \MathYTDXXVI{} list entries, \MathYTDOY\% above 2025 and \SCMYTDGap\% above a synthetic counterfactual of \SCMYTDSynthetic{} entries. Qualified donor and preperiod designs yield \SCMSensitivityLow\% to \SCMSensitivityHigh\%, and Mathematics has the largest RMSPE ratio among fifteen eligible placebo archives. Subfield growth is broad, with \SubfieldPositiveN{} of \SubfieldTotalN{} primary \code{math.*} categories expanding. The author panel shows a marked thickening of the repeated output tail. The share of active author units producing at least five submissions rose from \AuthorFivePlusActiveShareXXV\% to \AuthorFivePlusActiveShareXXVI\%, while the ten submission tail rose from \AuthorTenPlusActiveShareXXV\% to \AuthorTenPlusActiveShareXXVI\%. These results document a new and unusually large 2026 Mathematics production regime shift. Its timing and production structure, combined with independent evidence on AI diffusion and verifiable research tasks, are consistent with delayed diffusion and capability threshold mechanisms. The comparative design identifies the anomaly, and separate triangulation evaluates AI related explanations. The findings locate verification, selection, and attention as central constraints for research governance.
\end{abstract}


\section{Introduction}

The most visible scientific use of a \LLM{} is textual, yet its deepest effects reach the research process itself. A model can translate a question into code, propose a lemma, search a literature, produce a symbolic derivation, construct a numerical experiment, and turn intermediate results into a manuscript. These functions lower several production costs at once. Their scientific value rises sharply when outputs can be checked and the resulting questions remain worth asking. This advantage is especially consequential in mathematics and theoretical or computational physics, where formal proofs, executable code, limiting cases, conservation laws, and reproducible environments provide powerful validators. Broad syntheses now organize AI assisted discovery around hypothesis generation, experimental design, simulation, and interpretation, while recent work on scientific practice emphasizes task specific evaluation and human responsibility \parencite{wang2023discovery,binz2025practice}. In this study, a research gold rush denotes a rapid and broad expansion of entry and repeated output as production costs fall and perceived research opportunities widen.

The empirical debate has moved faster than its measurement. Global article production grew well before the public release of ChatGPT, with the estimated annual total rising from roughly 1.92 million articles in 2016 to 2.82 million in 2022 \parencite{hanson2024strain}. Bibliometric mapping also finds widespread growth in the direct use and potential benefits of AI across science since 2015, together with marked disciplinary and demographic disparities \parencite{gao2024quantifying}. A credible assessment therefore benefits from distinct measures of information flow, field exposure, linguistic influence, disclosed adoption, substantive task use, and knowledge value. Their joint movement supplies a stronger mechanism test than any single undifferentiated ``AI share.''

This paper uses the 2026 mathematics preprint surge as an empirical anchor for studying how generative AI is changing theoretical and computational research. The surge is treated as both a substantive phenomenon and a comparative identification problem. The empirical contribution is a monthly panel covering Mathematics, thirteen physics related archives, and six additional comparison archives, together with official global submission totals. The methodological contribution is a ladder of counterfactual tests comprising year over year comparisons, seasonal historical extrapolation, a regularized convex \SCM{}, spatial placebos, prior year pseudo holdouts, alternative donor pools, alternative preperiods, and leave one donor out refits. The conceptual contribution connects changing information flow to the emerging verification bottleneck and to human agency in research. A capability threshold denotes the point at which model performance and available checks make a research task reliable enough for routine workflow adoption.

Three findings organize the paper. First, the 2026 surge is a major platform information flow event. Mathematics had \MathYTDXXVI{} list entries through August, a \MathYTDOY\% year over year rise; broad physics and unique global submissions rose by \PhysicsYTDOY\% and \GlobalYTDOY\%, respectively. Second, the counterfactual ladder separates platform acceleration from Mathematics specific divergence. Mathematics stands \MathExcess\% above continuation of its own pre-2022 trend and \SCMYTDGap\% above the preferred weighted 2026 counterfactual. Qualified alternative donor and preperiod designs give \SCMSensitivityLow\% to \SCMSensitivityHigh\%. Third, the controlled divergence emerges in 2026 after small fitted gaps through 2025. This timing is consistent with delayed diffusion, accumulated workflow expertise, and a capability threshold rather than an immediate one time response to the public release of ChatGPT.

The author panel reveals a concurrent change in production structure. Single author submissions rose 54.7\% from 2025, author units with at least five submissions rose 81.6\%, and author units with at least ten submissions rose 170.1\%. These increases remain pronounced after normalizing by the expanding active author population. The 2026 surge therefore combines broader entry with a pronounced thickening of the repeated output tail.

The central estimand is the January--August 2026 divergence of the Mathematics archive list relative to a weighted set of other archive lists fitted through December 2025. This comparative quantity absorbs the broad evolution shared by donor fields and estimates the Mathematics specific anomaly that remains in the holdout period. December 2022 serves only as a public availability marker for generative AI. Independent population, author composition, subfield, and capability evidence then evaluates whether AI diffusion has the timing and task structure expected under the proposed mechanisms. In short, the synthetic control identifies the 2026 anomaly and triangulation evaluates AI related mechanisms.

Sections~\ref{sec:framework}--\ref{sec:methods} define the estimands, data, and methods. Sections~\ref{sec:results}--\ref{sec:capabilities} present the empirical results and mechanism evidence. Sections~\ref{sec:publishing}--\ref{sec:governance} examine publishing, human agency, and institutional responses. Sections~\ref{sec:discussion} and~\ref{sec:limitations} give falsifiable interpretations and study boundaries.

\section{Conceptual framework and estimands}
\label{sec:framework}

\subsection{Distinct empirical objects}

The phrase ``AI is producing more papers'' can refer to at least four different objects. They require different data and support different conclusions.

\begin{table*}[!t]
\centering
\caption{Measurement hierarchy for AI and scientific output}
\label{tab:measurement}
\begin{threeparttable}
\begin{tabularx}{\textwidth}{>{\raggedright\arraybackslash}p{2.8cm}Y Y Y}
\toprule
Object & Observable measure & What it supports & Interpretive scope \\
\midrule
Information flow & Archive list entries, unique submissions, versions & Workload, visibility, and platform pressure & Field exposure including cross lists \\
AI shaped language & Corpus level lexical mixture or excess vocabulary & Population level diffusion of stylistic influence & Population inference from textual distributions \\
Disclosed AI use & Author statements or manuscript classification & Observable adoption under current norms & Disclosure based adoption under evolving norms \\
Substantive AI contribution & Reported proof, formalization, coding, analysis, or problem formulation & Scientific task involvement & Task involvement paired with separate validation \\
Knowledge value & Correctness, novelty, robustness, reuse, and explanatory gain & Scientific contribution & Longitudinal and field specific evaluation \\
\bottomrule
\end{tabularx}
\begin{tablenotes}\footnotesize
\item The rows are not stages of one estimator. They are distinct constructs that may move at different rates.
\end{tablenotes}
\end{threeparttable}
\end{table*}

The first row is our directly measured outcome. Archive list entries represent what readers, moderators, and field communities encounter. They are therefore consequential measures of field exposure even when cross listing separates them from unique paper counts. The remaining rows provide triangulation evidence for mechanisms and governance.

\subsection{A causal graph and the target of the empirical design}

\begin{figure*}[!t]
\centering
\includegraphics[width=\textwidth]{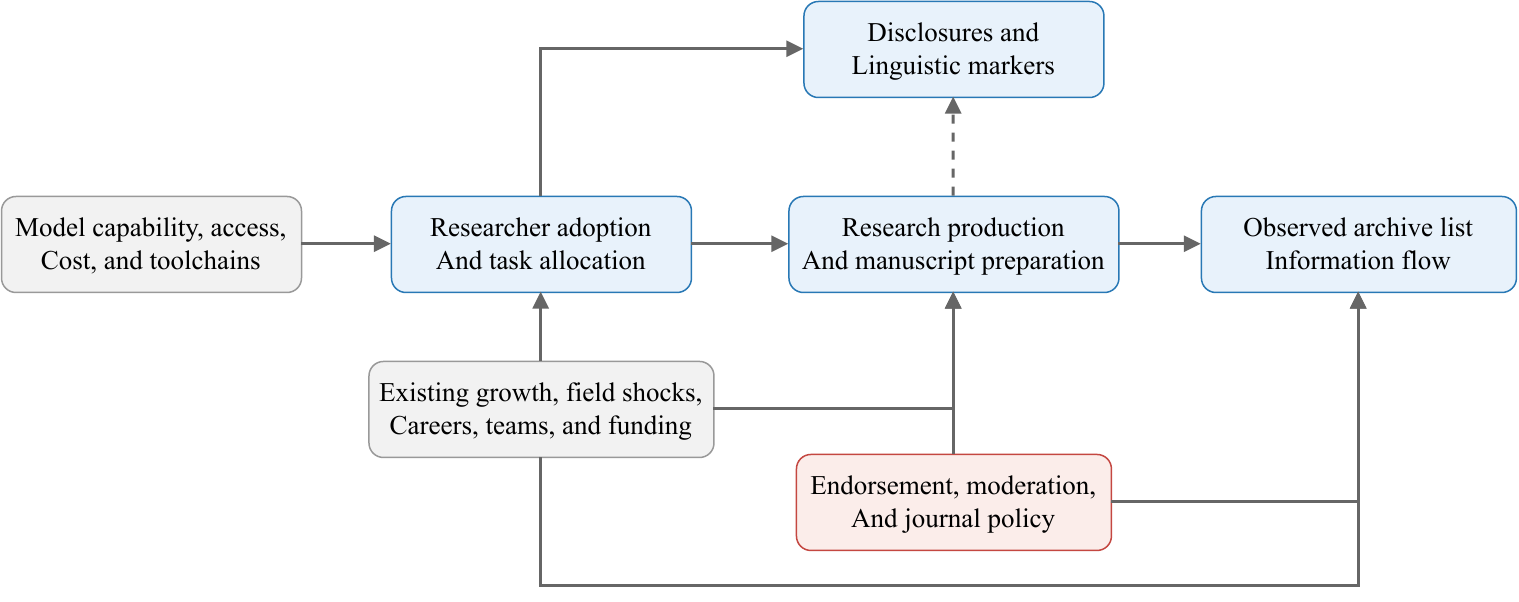}
\caption{Causal structure motivating the analysis. The synthetic control uses pre-2026 weighted donor trajectories to identify the 2026 Mathematics anomaly. Disclosures, linguistic markers, author structure, subfields, and capability evidence separately evaluate AI related mechanisms.}
\label{fig:dag}
\end{figure*}

Figure~\ref{fig:dag} summarizes the mechanisms connecting model capability to scientific information flow. Causal diagrams make the assumed pathways and adjustment target explicit \parencite{pearl1995causal}. Model capability and access influence researcher adoption; adoption changes research production and manuscript preparation; both affect submissions. Existing growth, field shocks, labor market incentives, collaboration, and platform policies shape the same pathway. Disclosure and language markers provide downstream measurements of adoption through separate measurement processes.

Using potential outcome notation \parencite{rubin1974estimating}, let $Y_{Mt}$ denote the seasonally adjusted log count for the Mathematics archive list in month $t$, and let $Y_{Mt}(0)$ denote the unobserved 2026 path that follows the weighted donor archive evolution learned from 2018--2025. Our comparative estimand for the January--August holdout period $T$ is
\begin{equation}
\Delta_T^{\mathrm{rel}}
=
\frac{\sum_{t\in T} N_{Mt}}
{\sum_{t\in T} \widehat{N}_{Mt}(0)}-1,
\label{eq:estimand}
\end{equation}
where $N$ is the count scale and $\widehat{N}_{Mt}(0)$ is the synthetic counterfactual. The estimand identifies the 2026 Mathematics list anomaly relative to the evolving donor archive system. We call it a comparative counterfactual because it quantifies the field specific component that remains after common archive dynamics are absorbed.

\section{Data and evidence architecture}
\label{sec:data}

\subsection{Official arXiv monthly counts}

The data window is January 2018 through August 2026. We queried twenty top level archives comprising \code{math}, thirteen physics related archives (\code{astro-ph}, \code{cond-mat}, \code{gr-qc}, \code{hep-ex}, \code{hep-lat}, \code{hep-ph}, \code{hep-th}, \code{math-ph}, \code{nlin}, \code{nucl-ex}, \code{nucl-th}, \code{physics}, and \code{quant-ph}), and six additional comparison archives (\code{cs}, \code{econ}, \code{eess}, \code{q-bio}, \code{q-fin}, and \code{stat}). The balanced panel contains $20\times104=2{,}080$ archive month observations.

Each archive page reports ``Total of $N$ entries.'' The list can include manuscripts listed from another primary archive \parencite{arxivMathList2026}. Within one archive month, it measures exposures or records processed by that field list. Summing archives can count a manuscript in multiple lists. The broad physics sum therefore approximates physics facing information and moderation load.

The official global monthly statistics provide unique submissions \parencite{arxivMonthlyStats}. We use that series for descriptive context and reserve the archive panel for synthetic control because the global total mechanically contains the target and donor archives. A limited primary category audit checks the list semantics and is reported in Appendix~\ref{app:archives}, while the balanced historical design uses the consistently available archive list series.

\begin{figure*}[!t]
\centering
\includegraphics[width=\textwidth]{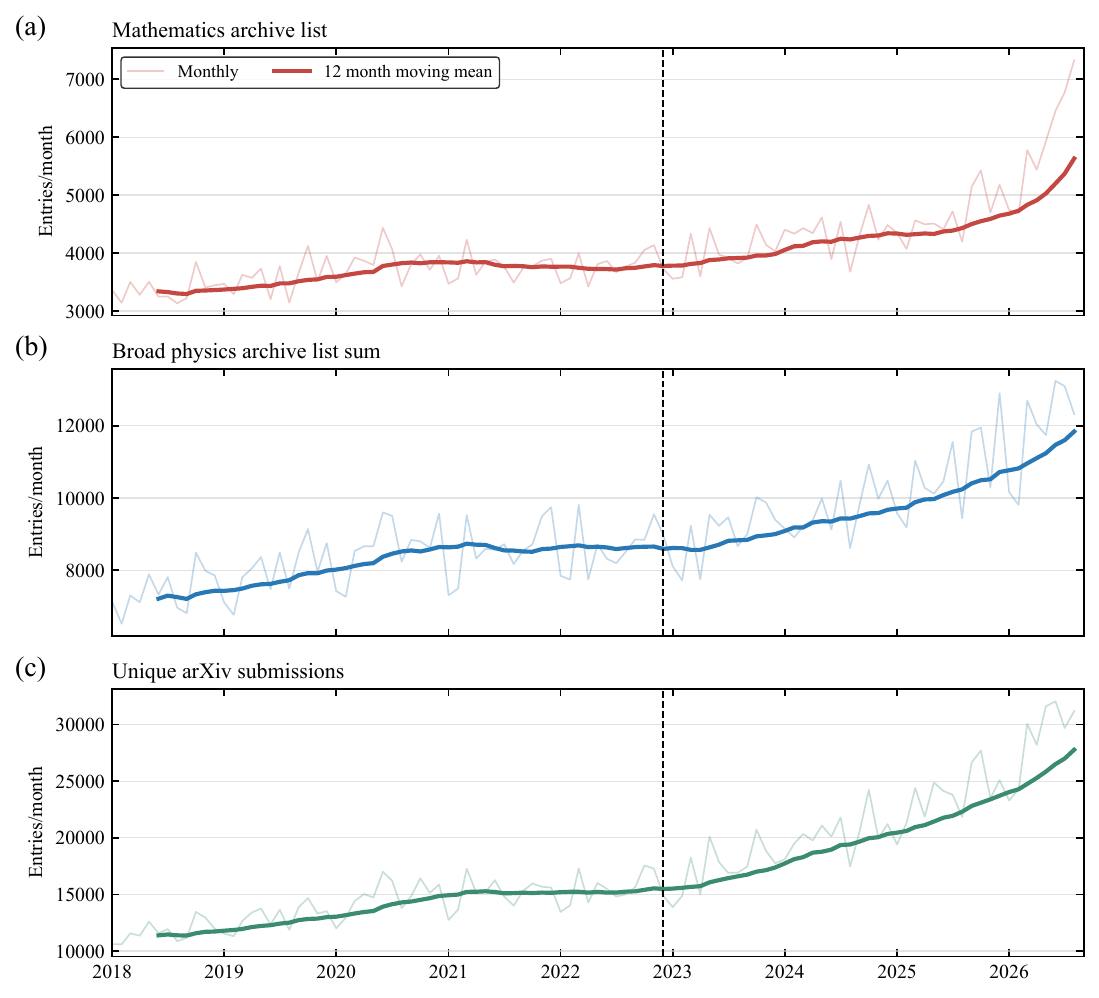}
\caption{Monthly arXiv information flows. Panels (a), (b), and (c) show the Mathematics archive list, the broad physics archive list sum, and unique platform submissions. Thin lines give monthly values and thick lines give 12 month moving means. Archive list counts include cross lists, so the physics sum can count one manuscript in multiple lists; the global series counts unique submissions. The dashed line marks December 2022 as a public availability reference for generative AI.}
\label{fig:trends}
\end{figure*}

\subsection{Pseudonymized author output panel}

We retrieved every unique first submission with at least one \code{math.*} category in January--August of 2023 through 2026 from the official Atom API \parencite{arxivAPI2026}. The API \code{published} field supplies the first submission date and ordered \code{author} elements supply author counts. Author display names were Unicode normalized and converted immediately to truncated SHA256 digests. The stored panel contains only pseudonymized author and paper digests, author position, author count, month, and category metadata.

For each matched eight month window, author output is the number of distinct participating submissions. High output denotes at least five submissions. Prior activity is the total in the two preceding matched January--August windows, grouped as none, one or two, and at least three. Exact normalized display name matching defines the longitudinal unit. Name variants remain separate and exact homonyms can merge, so the estimates characterize platform visible display name units. Definitions, privacy controls, and the full annual table appear in Appendix~\ref{app:authors}.

\subsection{Literature corpus and evidence standards}

Mechanism evidence comes from peer reviewed articles, preprints, and official platform and policy records on AI assisted writing, mathematics adoption, scientific capability, publishing pressure, and researcher or collective outcomes. Claims about new models, policies, and 2026 statistics are tied directly to their reporting sources.

This architecture turns measurement differences into an inferential advantage. Corpus language, author disclosure, substantive task involvement, archive volume, and later knowledge value arise through different measurement processes. Compatible movement across them strengthens mechanism inference while preserving a clear denominator for every reported percentage.

\section{Empirical strategy}
\label{sec:methods}

\subsection{Descriptive comparisons and seasonal historical extrapolation}

For each focal series we first report January--August 2026 counts and their change relative to January--August 2025. We then fit a log linear historical trend to January 2018--November 2022,
\begin{equation}
\log N_t=\alpha+\beta t+\sum_{m=2}^{12}\gamma_m\ind\{\mathrm{month}(t)=m\}+\varepsilon_t.
\label{eq:pretrend}
\end{equation}
The fitted model is extrapolated through August 2026. Confidence bands quantify uncertainty in the conditional historical mean. We also fit an interrupted time series specification with a post indicator and postperiod slope, using heteroskedasticity and autocorrelation consistent standard errors with twelve lags \parencite{neweywest1987simple}. These regressions establish each series' departure from its own history, while the synthetic control identifies comparative divergence across archives.

The historical model supplies a transparent baseline for the widely used ``excess above trend'' calculation. The comparative synthetic control is the main design. Changing the preperiod start from 2018 to 2020 moves the mathematics estimate from \MathExcessLow\% to \MathExcessHigh\%, which quantifies the leverage that modest slope differences acquire over several years.

\subsection{Regularized convex synthetic control}

Synthetic control constructs a weighted combination of comparison units that reproduces a target series before a held out period \parencite{abadie2003economic,abadie2010synthetic,abadie2021using}. January 2018 through December 2025 forms the preperiod, and January--August 2026 is held out from fitting. We first remove seasonality from log counts using calendar month means estimated exclusively in the preperiod. Let $Y_{it}^{*}$ be the resulting series for archive $i$. For the target $M$ and donor set $\mathcal{D}$, weights solve
\begin{equation}
\begin{aligned}
\widehat{\boldsymbol w}_{\lambda}
=\arg\min_{\boldsymbol w}\;&
\frac{1}{T_0}\sum_{t<T_H}
\left(Y_{Mt}^{*}-\sum_{j\in\mathcal{D}}w_jY_{jt}^{*}\right)^2 \\
&+\lambda\lVert\boldsymbol w\rVert_2^2, \\
\text{subject to}\;&w_j\geq0,\qquad \sum_jw_j=1.
\end{aligned}
\label{eq:scm}
\end{equation}
where $T_H$ is January 2026. The ridge term follows the regularization principle introduced for correlated predictors \parencite{hoerlkennard1970ridge}. It reduces unstable concentration while preserving nonnegative convex weights. We select $\lambda$ from a fixed grid by expanding window validation with 12 month validation blocks. The selected value is $0.01$; its validation mean squared error is $1.54\times10^{-3}$, compared with $1.78\times10^{-3}$ for the unregularized fit.

The primary donor set contains eighteen archives comprising all observed archives except Mathematics and \code{math-ph}. Conceptual overlap and cross listing connect \code{math-ph} directly to the target, so it enters a sensitivity analysis. The global total is reserved for descriptive context. Back transformation to counts uses the preperiod Mathematics seasonal factors. The primary preperiod log RMSE is \SCMPreRMSE{} and the effective donor count, $1/\sum_jw_j^2$, is 12.09.

Augmented synthetic control can combine balancing weights with an outcome model to correct residual preperiod imbalance \parencite{benmichael2021augmented}; Bayesian structural time series models provide a probabilistic alternative \parencite{brodersen2015causal}. We retain the convex baseline because it is transparent and the prefit is already close. These alternatives provide useful extensions for richer covariates and a longer holdout period.

\subsection{Design diagnostics}

We use spatial placebos, prior year pseudo holdouts, donor pool and preperiod sensitivity, and leave one donor out refits recommended in the synthetic control literature \parencite{abadie2021using}. Placebo archives with preperiod RMSE above five times the Mathematics value are screened before empirical ranking. For sensitivity analysis, a candidate specification $s$ with preperiod starting in year $y_0$ is classified as qualified when
\begin{equation}
\mathrm{RMSE}^{\mathrm{pre}}_{s,y_0}
\leq 2\,\mathrm{RMSE}^{\mathrm{pre}}_{\mathrm{primary},y_0}.
\label{eq:qualified-fit}
\end{equation}
This prespecified relative fit rule excludes counterfactuals that do not adequately reproduce Mathematics before the 2026 holdout and keeps the reported sensitivity range tied to credible preperiod comparability. Together the diagnostics establish exceptionalness relative to other archives, temporal specificity, donor and preperiod stability, and resistance to individual donor leverage. Full implementation details appear in Appendices~\ref{app:scm} and~\ref{app:weights}.

\section{Results of the 2026 divergence}
\label{sec:results}

\subsection{Raw changes and the choice of denominator}

Table~\ref{tab:headline} reports the main descriptive changes. Mathematics increased from 35,307 list entries in January--August 2025 to \MathYTDXXVI{} in the same months of 2026, a \MathYTDOY\% increase. The broad physics list sum rose \PhysicsYTDOY\%, while unique global submissions rose \GlobalYTDOY\%. The global rise establishes a platform wide acceleration, and the much larger Mathematics increase establishes a field specific amplification within it.

\begin{table}[!t]
\centering
\caption{January--August arXiv volume under three measurement conventions}
\label{tab:headline}
\begin{threeparttable}
\footnotesize
\setlength{\tabcolsep}{3pt}
\begin{tabularx}{\columnwidth}{Y r r r}
\toprule
Series & 2025 & 2026 & Change \\
\midrule
Mathematics archive list & 35,307 & \MathYTDXXVI{} & \MathYTDOY\% \\
Broad physics archive list sum & 81,647 & \PhysicsYTDXXVI{} & \PhysicsYTDOY\% \\
Unique arXiv submissions & 181,595 & \GlobalYTDXXVI{} & \GlobalYTDOY\% \\
\bottomrule
\end{tabularx}
\begin{tablenotes}\footnotesize
\item Archive list counts measure field exposure and can include cross lists. The broad physics sum can count the same paper in multiple archives. The global series counts unique submissions.
\end{tablenotes}
\end{threeparttable}
\end{table}

Growth within physics is heterogeneous. The largest absolute 2026 increases occur in quantum physics (+3,026; +30.2\%), condensed matter (+2,470; +15.8\%), astronomy and astrophysics (+2,014; +14.7\%), the general physics archive (+1,839; +11.1\%), high energy theory (+1,030; +20.3\%), and mathematical physics (+1,019; +31.1\%). Nuclear experiment entries rise 2.6\%, while electrical engineering and systems decline 8.6\% outside the physics sum. This heterogeneity motivates archive level weights.

The single series extrapolation assigns excesses of \MathExcess\% for Mathematics, \PhysicsExcess\% for broad physics, and \GlobalExcess\% for unique global submissions. Their joint rise quantifies the platform wide acceleration, while the comparative design below extracts the additional Mathematics divergence.

\subsection{The comparative counterfactual}

\begin{figure*}[!t]
\centering
\includegraphics[width=0.91\textwidth]{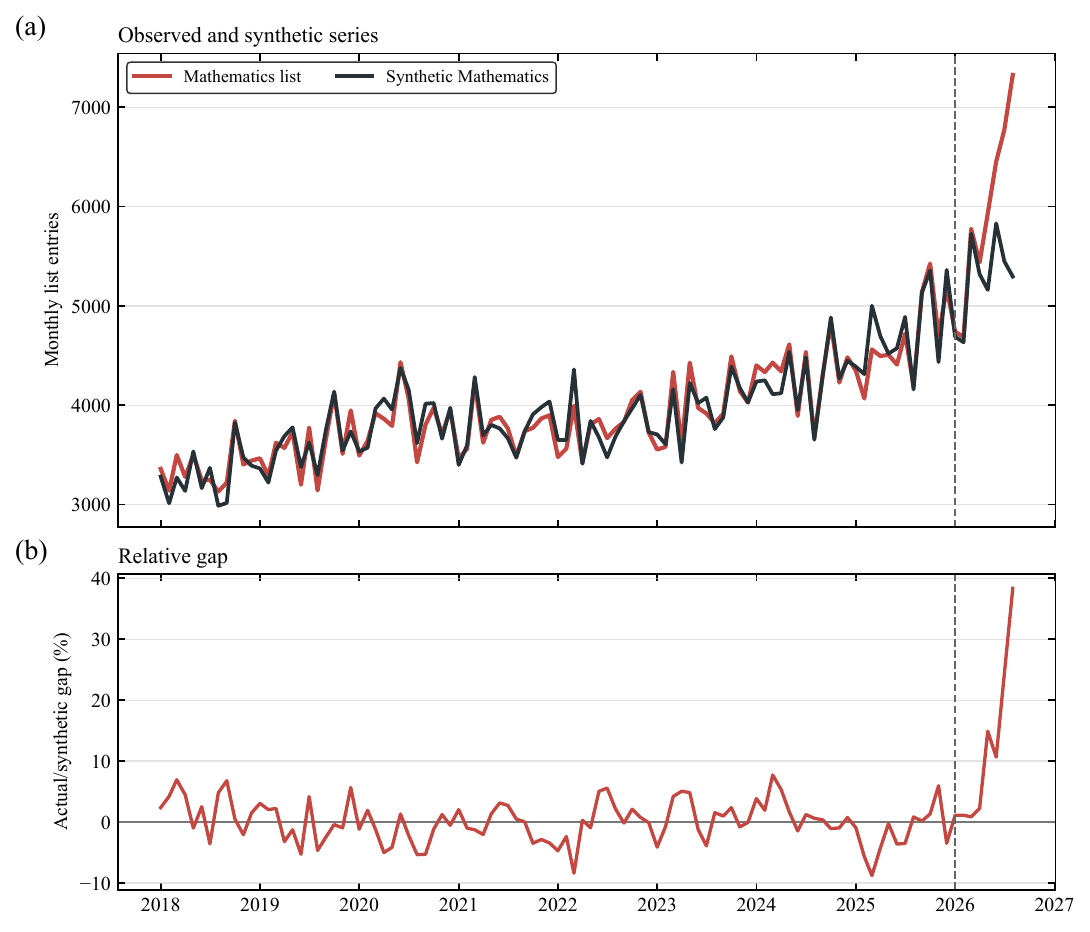}
\caption{Comparative counterfactual for Mathematics. Panel (a) shows observed archive list entries and the seasonally adjusted convex synthetic control. Panel (b) shows the monthly observed to synthetic gap. The dashed line marks the January 2026 holdout boundary. The subsequent separation quantifies the Mathematics specific component beyond the weighted evolution of donor archives.}
\label{fig:scm}
\end{figure*}

Figure~\ref{fig:scm} shows the preferred synthetic control. The monthly fit is close through December 2025, after which January--August 2026 is evaluated as a genuine holdout. Actual Mathematics has \SCMYTDActual{} entries and synthetic Mathematics has \SCMYTDSynthetic{} in the holdout, implying a \SCMYTDGap\% relative gap.

\begin{table}[!t]
\centering
\caption{Actual and synthetic Mathematics around the 2026 holdout}
\label{tab:annualgap}
\footnotesize
\setlength{\tabcolsep}{3pt}
\begin{tabularx}{\columnwidth}{Y r r r}
\toprule
Period & Actual & Synthetic & Gap \\
\midrule
2022 & 45,310 & 45,399 & \SCMGapXXII\% \\
2023 & 47,789 & 47,462 & \SCMGapXXIII\% \\
2024 & 52,081 & 51,248 & \SCMGapXXIV\% \\
2025 & 55,753 & 56,834 & \SCMGapXXV\% \\
January--August 2026 & \SCMYTDActual{} & \SCMYTDSynthetic{} & \SCMGapXXVI\% \\
\bottomrule
\end{tabularx}
\end{table}

The fitted annual gaps are \SCMGapXXII\% in 2022, \SCMGapXXIII\% in 2023, \SCMGapXXIV\% in 2024, and \SCMGapXXV\% in 2025, followed by \SCMGapXXVI\% in the 2026 holdout. The break is therefore concentrated in 2026 rather than spread across the years following public access to ChatGPT. This timing is consistent with delayed diffusion and capability threshold mechanisms. Accumulated workflow expertise, changing author composition, submission behavior, policy anticipation, and topical shocks supply additional testable channels for the amplification.

\subsection{Placebos and robustness}

In the spatial placebo exercise, Mathematics has the largest holdout to preperiod RMSPE ratio among the \SCMPlaceboN{} archives meeting the fit criterion. The finite sample empirical placebo probability is $1/15=\SCMPlaceboP$. The rank establishes Mathematics as the strongest proportional break in the eligible archive system.

\begin{figure}[!t]
\centering
\includegraphics[width=\columnwidth]{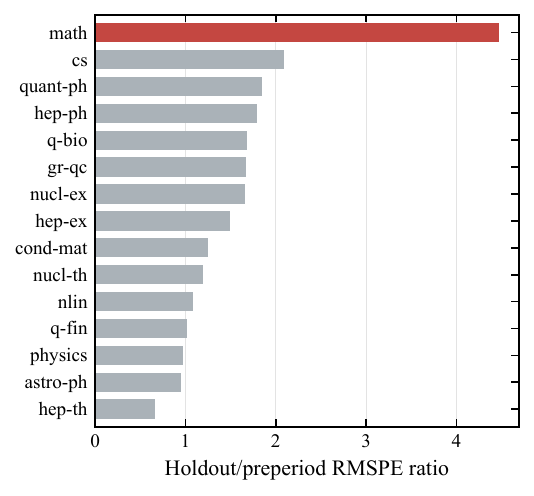}
\caption{Spatial placebo distribution. Four archives with preperiod RMSE above five times the Mathematics value are screened, leaving 15 units that meet the fit criterion. Mathematics is highlighted and ranks first. The empirical probability includes the target archive under the finite placebo ranking convention.}
\label{fig:placebo}
\end{figure}

Prior year pseudo holdouts reinforce the temporal specificity of the result. Reestimating the model before each matched January--August window gives gaps of \SCMPseudoXXIII\% in 2023, \SCMPseudoXXIV\% in 2024, and \SCMPseudoXXV\% in 2025, compared with \SCMPseudoXXVI\% in 2026. Across every donor pool and 2018, 2019, or 2020 preperiod start that meets the fit criterion, the 2026 estimate ranges from \SCMSensitivityLow\% to \SCMSensitivityHigh\%. The six nonphysics specifications have prefit RMSE more than twice the corresponding primary fit and are shown as hollow markers outside the qualified set. Thus the 2026 anomaly persists when early preperiod years are removed as well as when donor composition changes.

\begin{figure*}[!t]
\centering
\includegraphics[width=0.96\textwidth]{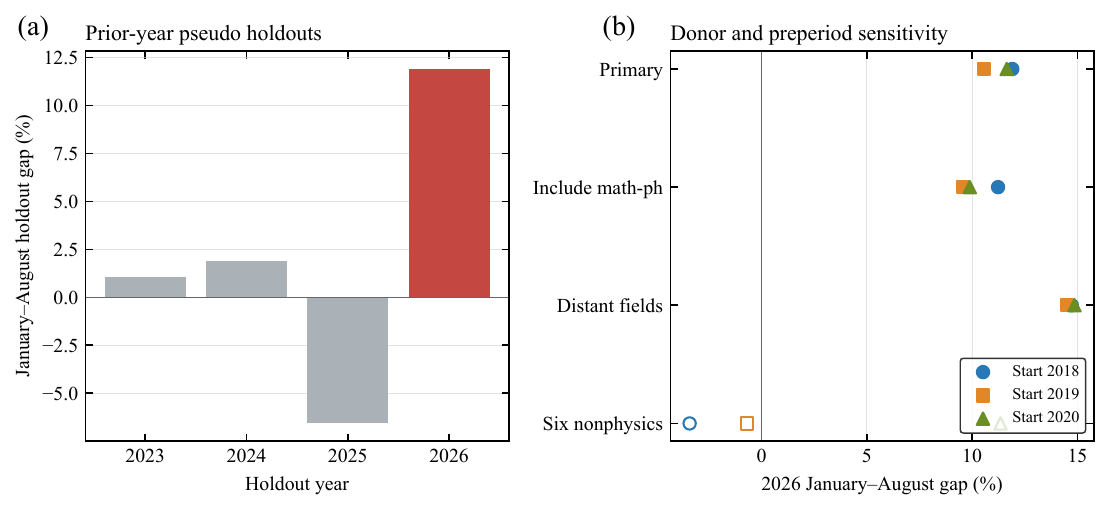}
\caption{Temporal, donor pool, and preperiod robustness. Panel (a) gives January--August gaps from prior year pseudo holdouts and the 2026 holdout. Panel (b) gives 2026 estimates for four donor pools with preperiods beginning in 2018, 2019, or 2020. Filled markers meet the prespecified fit threshold; hollow markers do not.}
\label{fig:robustness}
\end{figure*}

Leave one donor out refits place the 2026 gap between \SCMLeaveOneOutLow\% and \SCMLeaveOneOutHigh\%. This compact range demonstrates that the headline result is distributed across the donor combination rather than driven by one weighted archive.

\subsection{What the quantitative evidence licenses}

The evidence supports three cumulative claims.

\begin{enumerate}
\item \textbf{Volume result.} The Mathematics archive list experienced a sharp information flow increase in 2026, larger than its 2025 level and larger than broad physics growth.
\item \textbf{Comparative result.} Mathematics diverged positively from a weighted combination of other archive lists in 2026. The preferred magnitude is \SCMYTDGap\%, and qualified donor and preperiod alternatives range from \SCMSensitivityLow\% to \SCMSensitivityHigh\%.
\item \textbf{Mechanism evidence.} The late timing, population evidence of AI diffusion, broad subfield growth, a thicker repeated output tail, and expansion of verifiable model capabilities are jointly consistent with delayed AI diffusion and a capability threshold.
\end{enumerate}

The first two levels establish the scale and comparative exceptionalness of the divergence. The third triangulates its timing, production structure, and task capability. The synthetic control identifies the anomaly rather than the share caused by AI. Manuscript level linkage between disclosure, task use, version history, and validation outcomes offers the next direct estimate of the AI contribution.

\section{AI diffusion and author output dynamics}
\label{sec:adoption}

Three recent measurement strategies document rapid diffusion through distinct constructs. A mixture model applied to roughly 1.12 million papers estimated that, by September 2024, \LLM{} modification reached approximately 22\% in computer science and 9\% in a Mathematics and Nature grouping \parencite{liang2025usage}. Excess vocabulary analysis of 14 million PubMed abstracts placed a lower bound near 10\% in 2024, with higher estimates in some subcorpora \parencite{kobak2025delving}. A 2026 analysis of approximately 7.3 million full texts from four large publishers reported an increase in broadly defined LLM influenced text from about 12\% in 2023 to 57\% in 2025 \parencite{siler2026diffusion}. Together these population measures establish broad and rapidly increasing linguistic diffusion.

Official author metadata show that the 2026 volume change also altered the composition of mathematical production. Unique submissions with at least one Mathematics category increased from \AuthorSubmissionsXXV{} in January--August 2025 to \AuthorSubmissionsXXVI{} in the matched 2026 window, a 33.6\% rise. This API total is a distinct denominator from the archive list total. It counts unique first submissions carrying any Mathematics category, whereas the archive series follows regular list exposure. The close agreement between the 33.6\% unique submission increase and the 33.5\% list increase shows that the surge is present under both denominators. Active author units increased from \AuthorUniqueXXV{} to \AuthorUniqueXXVI{}, or 17.2\%, so submissions grew nearly twice as fast as participating units. Single author submissions rose from \AuthorSingleXXV{} to \AuthorSingleXXVI{}, a 54.7\% increase, and their share rose from \AuthorSingleShareXXV\% to \AuthorSingleShareXXVI\%.

\begin{figure*}[!t]
\centering
\includegraphics[width=0.98\textwidth]{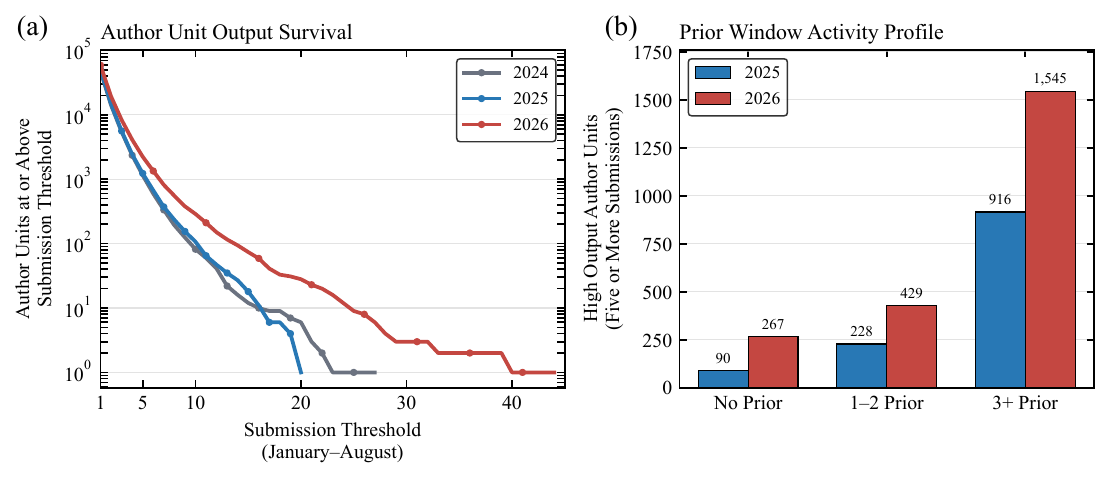}
\caption{Pseudonymized author output dynamics in official arXiv metadata \parencite{arxivAPI2026}. Panel (a) gives the number of author units at or above each submission threshold during January--August. Panel (b) classifies author units with at least five submissions by activity in the two preceding matched eight month windows. Each plotted unit is a pseudonymized exact normalized display name.}
\label{fig:authoroutput}
\end{figure*}

The repeated output tail thickened strongly. Author units with at least five submissions increased from \AuthorFivePlusXXV{} to \AuthorFivePlusXXVI{}, and units with at least ten increased from \AuthorTenPlusXXV{} to \AuthorTenPlusXXVI{}. The shift is not only a scale effect. The share of active author units producing at least five submissions rose from \AuthorFivePlusActiveShareXXV\% to \AuthorFivePlusActiveShareXXVI\%, while the ten submission tail more than doubled from \AuthorTenPlusActiveShareXXV\% to \AuthorTenPlusActiveShareXXVI\%. The maximum rose from \AuthorMaximumXXV{} to \AuthorMaximumXXVI{}. The author output Gini increased from \AuthorGiniXXV{} to \AuthorGiniXXVI{}, while the top one percent share of author paper participations increased from \AuthorTopOneShareXXV\% to \AuthorTopOneShareXXVI\%. We operationalize a mining like pattern as this aggregate combination of rapid entry, repeated production, and a thicker extreme tail. The label applies to the measured production configuration rather than individual identity or scientific merit.

Primary category decomposition shows that the surge is broad rather than a single topic shock. Among \SubfieldTotalN{} primary \code{math.*} categories, \SubfieldPositiveN{} grew between the matched 2025 and 2026 windows. \code{\SubfieldLargestCode} contributed the largest absolute increment with \SubfieldLargestIncrease{} additional submissions, or \SubfieldLargestShare\% of the total increase, and grew \SubfieldFastLargeGrowth\%, the fastest rate among categories with at least 1,000 submissions in 2025. The ten largest contributors account for \SubfieldTopTenShare\% of the increase. This combination of near universal direction and heterogeneous magnitude supports general diffusion with subfield specific amplification. Appendix Figure~\ref{fig:subfields} reports the full decomposition.

Historical activity profiles locate the source of the new tail. Among high output author units, the group with no matched activity in the preceding two windows increased from \AuthorHighNoPriorXXV{} in 2025 to \AuthorHighNoPriorXXVI{} in 2026. Its share rose from \AuthorHighNoPriorShareXXV\% to \AuthorHighNoPriorShareXXVI\%. The number with no more than two prior submissions increased from \AuthorHighLowPriorXXV{} to \AuthorHighLowPriorXXVI{}. These movements show that the 2026 expansion combines continued production by established platform visible units with a substantially larger inflow of units that had low prior window arXiv visibility.

Other observational evidence shows that output and measured quality can move separately. A study combining 2.1 million preprints, roughly 28,000 reviews, and readership data reports output increases of 23.7--89.3\% among inferred adopters across fields and backgrounds, together with more complex language and declines in several quality proxies \parencite{kusumegi2026production}. Its classifier and matching design support a strong association between adoption, greater production, and a redistribution of measured quality.

Population estimates of AI shaped text establish rapid diffusion across science. The author panel independently establishes a change in Mathematics production structure. Their temporal alignment with the late comparative divergence, together with the capability evidence in Section~\ref{sec:capabilities}, is consistent with a diffusion process in which broadly available tools lower entry and repetition costs after workflows mature. Linking stable arXiv identifiers, version histories, task level disclosures, and later verification outcomes can estimate the mediated contribution of documented AI assisted workflows.

\section{Capability frontiers in mathematics and theoretical physics}
\label{sec:capabilities}

\subsection{Verification changes the production function}

Research tasks vary in how cheaply their outputs can be checked. This difference helps explain why progress can be fast in some parts of mathematics and computation while open ended theory remains verification intensive. A candidate program can be executed; a formal theorem can be checked by a kernel; and a numerical result can be rerun when code and environments are preserved. Novelty, explanatory significance, a physical approximation's domain of validity, and literature completeness require deeper expert evaluation. Work on scientific understanding accordingly distinguishes successful prediction from explanations that expose mechanisms, representations, and relations that scientists can use \parencite{krenn2022understanding}.

The capability record illustrates this gradient. AI guided analysis of mathematical data helped human mathematicians form and prove conjectures in knot theory and representation theory \parencite{davies2021intuition}. FunSearch combined model generated programs with an automatic evaluator and evolutionary selection to produce new, checkable constructions in problems including cap sets \parencite{romeraparedes2024funsearch}. AlphaGeometry paired a neural language component with symbolic deduction and solved 25 of 30 olympiad geometry problems in its reported evaluation \parencite{trinh2024alphageometry}. DeepMind later reported a gold medal level score for a natural language system at the 2025 International Mathematical Olympiad \parencite{deepmindIMO2025}. The sequence traces a frontier moving from machine assisted pattern discovery toward complete, externally checked solutions.

Research level formalization provides a demanding frontier. RLMEval covers 613 theorems from six Lean projects and reports a best \code{pass@128} of 10.3\% for normal mode proof autoformalization \parencite{poiroux2025rlmeval}. The First Proof project tested four systems on ten previously unseen research problems in its second batch. Seven problems received at least one passing expert grade among the submitted solutions \parencite{abouzaid2026firstproof}. Problem secrecy, interaction logs, proof checking, conflicts of interest, and expert review criteria make these outcomes auditable. Together they show that increasingly open research tasks become tractable when generation is paired with structured verification.

\begin{figure*}[!t]
\centering
\includegraphics[width=\textwidth]{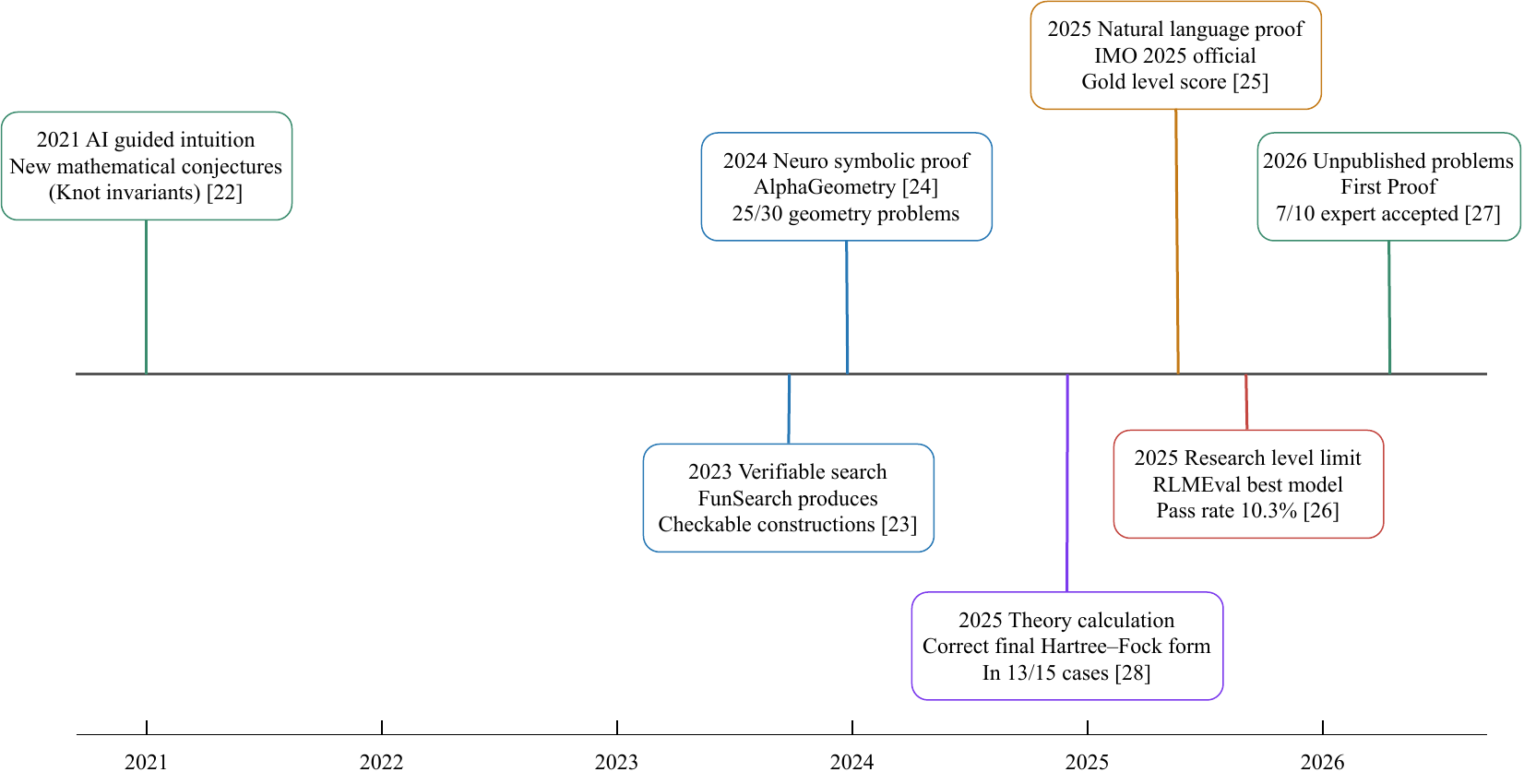}
\caption{Selected capability landmarks from mathematical intuition to expert reviewed solutions of unpublished problems \parencite{davies2021intuition,romeraparedes2024funsearch,trinh2024alphageometry,pan2025manybody,deepmindIMO2025,poiroux2025rlmeval,abouzaid2026firstproof}. The lower scale marks calendar years. Distinct task and validation regimes show how executable evaluation, symbolic checking, formal proof, and expert review progressively expand the frontier of verifiable machine contribution.}
\label{fig:capability}
\end{figure*}

The emerging mathematical workflow has a characteristic division of labor. Machines generate candidates, search combinatorial spaces, translate informal statements into formal ones, and attempt proof repair. Humans decide which representation is meaningful, identify missing assumptions, construct discriminating examples, connect a formal certificate to an explanatory argument, and judge why a theorem matters. Machine assistance can therefore increase human agency when it expands the set of testable options. It reduces agency when researchers accept an opaque proposal because producing an independent understanding is too expensive. Contemporary reflections on mathematics in the age of AI similarly emphasize coexistence among machine assistance, proof, explanation, and mathematical culture \parencite{tao2026ageai,klowden2026humanthought}.

\subsection{Theoretical and computational physics}

Physics already contains mature forms of machine assistance. Neural network quantum states represent interacting many body wave functions \parencite{carleo2017manybody}, and broad reviews document machine learning across physical modeling \parencite{carleo2019physics}. More recent reviews extend this landscape to physics informed learning and to data driven discovery, reduction, and solution operators for partial differential equations \parencite{karniadakis2021physicsinformed,brunton2024pde}. Symbolic regression systems such as AI Feynman search for compact formulas under structured constraints \parencite{udrescu2020aifeynman}. At materials scale, graph networks paired with density functional calculations have expanded the search space to millions of candidate crystals while retaining a computational verification layer \parencite{merchant2023materials}. \GenAI{} adds a natural language interface across symbolic manipulation, literature, and code, making physical validation an organizing element of the workflow.

In one structured study, stepwise prompting enabled GPT-4 to reconstruct Hartree--Fock models from fifteen published cases; thirteen final Hamiltonians were correct and average step scores were 87.5/100 \parencite{pan2025manybody}. Templates, intermediate representations, and correction loops produced the strongest performance. Language agent experiments in theoretical physics likewise combine retrieval, reasoning, code execution, and feedback into longer workflows, with error detection and benchmark coverage serving as central design requirements \parencite{lu2025physicsagents}.

Evidence retrieval is itself a scientific constraint. A high temperature superconductivity evaluation constructed a curated corpus of 1,726 papers and 67 expert questions. Curated retrieval augmented systems outperformed general closed models on key criteria \parencite{pan2026hightc}. Their remaining errors in figures, conflicting claims, and evidential sufficiency identify where expert judgment adds greatest value, including phase assignment, sample condition, pressure path, and experimental artifacts.

For computational physics, four validation layers organize the workflow.
\begin{enumerate}
\item \textbf{Formal validity.} Equations, units, symmetries, and code syntax.
\item \textbf{Numerical validity.} Convergence, conditioning, discretization, and stochastic uncertainty.
\item \textbf{Physical validity.} Limiting cases, conservation laws, parameter regimes, and comparison with observation.
\item \textbf{Epistemic validity.} Provenance, novelty, negative results, and competing explanations.
\end{enumerate}
Automation is strongest at the first layer and often useful at the second. The third and fourth remain deeply dependent on domain judgment and on evidence not fully represented in text corpora.

\section{How the publication mechanism changes}
\label{sec:publishing}

\subsection{From a production bottleneck to a verification bottleneck}

Traditional publishing evolved when preparing a coherent manuscript was expensive and therefore provided a weak signal of effort and selectivity. As drafting, translation, formatting, coding, and literature summarization become cheaper, that signal erodes. In a randomized experiment on bounded professional writing tasks, access to ChatGPT reduced completion time by 40\% and increased assessed output quality by 18\%, demonstrating the scale of the supply side shift that generative assistance can produce \parencite{noy2023productivity}. The scarce resources become expert attention, reliable replication, dataset and code inspection, and the capacity to decide which questions deserve community time.

This change creates a verification multiplier. If one researcher can produce $k$ times as many plausible manuscripts while reviewer capacity remains fixed, the average verification budget per manuscript falls roughly as $1/k$ unless institutions add screening or validation resources. The risk is not merely false papers. Large volumes of superficially adequate work can make important results harder to find, fragment claims across minimally distinct papers, and increase the cost of establishing priority and consensus.

The expansion of candidate supply is now technically concrete. An end to end system reported in 2026 generated research ideas, wrote and executed code, analyzed experiments, prepared manuscripts, and performed automated review in bounded machine learning settings \parencite{lu2026aiscientist}. Such systems make verification capacity a direct determinant of whether faster generation produces cumulative knowledge or additional screening load.

\arxiv{} provides a revealing case because moderated preprint dissemination precedes journal peer review. Bibliometric analysis of the complete archive and the Web of Science has established systematic timing and impact relationships between arXiv e-prints and their journal versions \parencite{lariviere2014arxiv}. The platform's current policy requires disclosure of significant generative text use, holds human authors responsible, rejects AI systems as authors, and permits restrictions on excessive submissions \parencite{arxivModeration2026}. In late 2025 and January 2026 the platform expanded endorsement requirements in response to an unsustainable increase in nonscientific submissions, higher rejection rates, and staff burden \parencite{arxivEndorsement2026}. The policy change supplies direct institutional evidence that intake and moderation capacity had become binding.

Endorsement can reduce abuse while increasing barriers for independent scholars, researchers in weakly connected institutions, and entrants from new fields. Platform response should therefore be evaluated through moderator time per accepted item, false acceptance and false rejection rates, author concentration, and appeal outcomes. This outcome set rewards scientific value and procedural fairness alongside screening volume.

\subsection{Peer review, disclosure, and credibility debt}

AI is also entering review. Corpus level estimates for four AI conferences found LLM modification in as much as approximately 17\% of some review sets \parencite{liang2024peerreview}. The immediate risks are confidentiality, fabricated critique, synchronized stylistic bias, and responsibility gaps. AI can assist a reviewer by locating an inconsistency or checking a calculation, while the reviewer retains a fully defensible judgment and auditable reasoning path.

A study of 5,114 journals and more than 5.2 million papers reports that roughly 70\% of sampled journals had an AI policy, while detected AI writing trends were similar in journals with and without such policies \parencite{he2026policies}. This evidence supports a decisive distinction between policy presence and operational transparency. Recent editorial guidance likewise identifies transparency, human accountability, fact checking, and review confidentiality as operational requirements for responsible use \parencite{naturemethods2026responsible}. Effective governance requires auditable disclosure, enforcement records, and measurable verification outcomes.

We use \emph{credibility debt} to describe unverifiable labor shifted downstream. An undisclosed generated derivation, an unpinned environment, an unchecked citation list, or an AI produced review can save minutes for the producer and impose hours on readers. Like technical debt, credibility debt compounds when later work relies on the unverified result. Publication systems should make that debt observable and assign its cost closer to the point of production.

\subsection{The paper as an expanded evidence unit}

The paper will remain useful for narrative, priority, and synthesis. Its minimum credible unit is likely to expand from prose plus static figures to a structured evidence package comprising claims, provenance, executable code, formal certificates where appropriate, robustness tests, model and prompt disclosure when scientifically material, and a human readable explanation of failure conditions.

This shift may separate functions now bundled in one article. A result can be rapidly registered as a preprint, formally certified later, replicated independently, and integrated into a living synthesis. Journals and platforms should preserve links among these objects and position the PDF within a larger durable record. Version history then becomes evidence rather than noise.

\section{Human agency and researcher adaptation}
\label{sec:agency}

Human agency extends beyond performing every operation manually. It is the capacity to select goals, understand constraints, contest outputs, change course, and remain responsible for consequences. A scientist exercises agency when AI expands the hypothesis set while the scientist explains the selection rule, reconstructs consequential inferences, knows what information entered the model context, and rejects attractive failures. Human aware models that incorporate the distribution of expertise can improve predictions of future discoveries and search for valuable hypotheses outside crowded trajectories, providing a concrete design route for complementarity \parencite{jia2023humanaware}.

The distinction matters because individual and collective incentives can diverge. Large scale observational work reports that AI augmented researchers are associated with 3.02 times as many annual papers, 4.84 times as many total citations, and becoming principal investigators 1.37 years earlier, while scientific topic space contracts 4.63\%, follow on engagement declines 22\%, and average team size is 1.33 researchers smaller \parencite{hao2026impact}. The matched and modeled estimates reveal a strong pattern of private productivity gains coexisting with public knowledge externalities.

\begin{figure*}[!t]
\centering
\includegraphics[width=\textwidth]{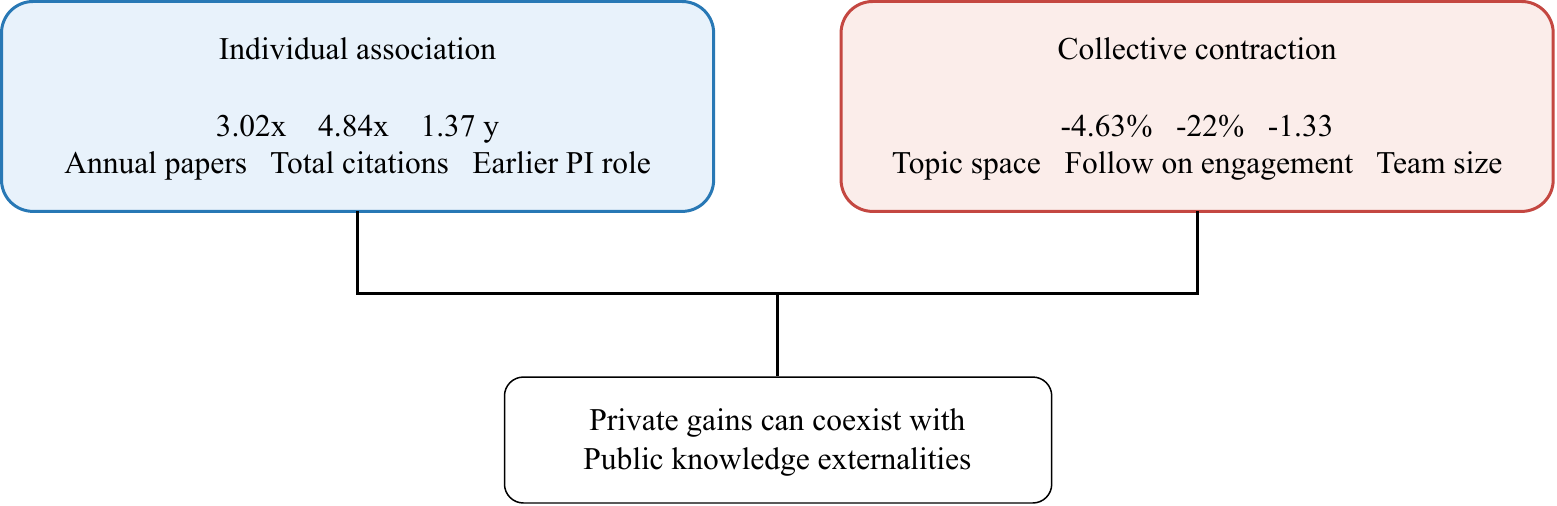}
\caption{The individual and collective pattern estimated from 41.3 million natural science papers \parencite{hao2026impact}. The left card reports matched individual associations with output, citations, and earlier principal investigator status. The right card reports model based changes in topic space, follow on engagement, and team size. Together the estimates support coexistence between private productivity gains and collective knowledge externalities.}
\label{fig:paradox}
\end{figure*}

One plausible mechanism is model mediated convergence. Models interpolate most effectively within abundant, digitized, and well evaluated domains. Adopters can enter those domains efficiently, while questions requiring new instruments, unusual data, tacit knowledge, or long periods without publishable output become relatively less attractive. The result can be more work per person alongside less exploration by the collective. A randomized writing experiment exhibits the same micro level pattern: generative AI improved individual evaluations while making outputs more similar to one another \parencite{doshi2024creativity}. The ``illusions of understanding'' framework similarly shows how universal productivity, objectivity, and understanding narratives can reinforce monocultures and conceal epistemic limits \parencite{messeri2024illusions}.

Researchers can respond with a task ledger that records the delegated task, model, retrieval corpus, toolchain, information sent, output retained, independent check, and responsible human decision. The ledger scales with scientific risk. Translating a paragraph requires less detail than generating a theorem statement, choosing a physical approximation, or modifying analysis code.

A practical six checkpoint protocol combines question ownership, an explicit assumption inventory, adversarial generation, independent verification, provenance, and risk calibrated disclosure. The operational details are placed in Appendix~\ref{app:protocol} so that the main argument remains focused on measured institutional change.

Survey evidence from nearly 5,000 researchers across more than 70 countries suggests that AI assistance in manuscripts, grants, and review will become more accepted while demand for training and institutional support remains high \parencite{naddaf2025survey}. The adaptation problem is therefore organizational as well as individual. Early career researchers need opportunities to learn derivations, debugging, and critical reading alongside productivity. Senior researchers need incentives to document negative checks and mentor conceptual understanding.

\section{Institutional design and a monitoring dashboard}
\label{sec:governance}

\subsection{Risk based use}

Policies should regulate scientific risk and information flow across changing product names. ICMJE and Nature Portfolio policies place authorship responsibility on humans, require transparent use and content verification, and protect confidential review material \parencite{icmjeAI2026,natureAIPolicy2026,naturemethods2026responsible}. A practical implementation is shown in Table~\ref{tab:risk}.

\begin{table*}[!t]
\centering
\caption{A risk based framework for AI assisted research}
\label{tab:risk}
\begin{tabularx}{\textwidth}{p{2.25cm}Y Y Y}
\toprule
Tier & Examples & Minimum control & Publication disclosure \\
\midrule
Low & Grammar, formatting, nonsubstantive translation & Human reading; no confidential data & Brief tool and purpose statement where required \\
Moderate & Literature triage, code completion, algebra, plotting, outline generation & Source verification; unit tests; environment capture; sampling of rejected output & Task level description and material model or tool versions \\
High & Proof construction, model selection, physical interpretation, data exclusion, reviewer recommendation & Independent method or expert; full provenance; adversarial tests; conflict and confidentiality review & Detailed contribution and validation record \\
Unacceptable without redesign & Invented citations, unreviewed autonomous decisions, upload of confidential manuscripts to unauthorized systems & Stop workflow; contain and report exposure; repeat analysis from trusted inputs & Correction or incident disclosure as applicable \\
\bottomrule
\end{tabularx}
\end{table*}

\subsection{Budget verification explicitly}

Funding and project plans should include a verification budget alongside compute and personnel. Relevant line items include formalization, replication, independent code review, benchmark creation, expert reading of figures and source data, and maintenance of curated corpora. Total research cost includes both generation and verification. Review panels should reward reusable validators, negative evidence, and high value manuscripts.

Platforms can use staged friction. New or unusually high volume submitters may face enhanced provenance checks, while quality controls apply consistently across career stages. Automated triage should be monitored for field, language, institution, and country disparities. Appeals need a human path. Rate limits should be paired with mechanisms for legitimate large collaborations or linked result packages.

\subsection{A quantitative dashboard}

The dashboard in Table~\ref{tab:dashboard} separates volume, process, quality, diversity, and distribution. Its indicators also supply outcomes for policy evaluation. If an endorsement rule is introduced at a known date, a controlled interrupted design can compare affected domains and jointly examine rejected volume, accepted quality, moderator load, appeals, and representation. The complete indicator set is reported in Appendix~\ref{app:dashboard}.

\section{Discussion and falsifiable mechanisms}
\label{sec:discussion}

\subsection{Mechanisms for the 2026 divergence}

At least five mechanisms are consistent with the timing.

\begin{enumerate}
\item \textbf{Delayed capability threshold.} Improvements in long context reasoning, tool use, formal systems, and coding made substantive mathematical assistance viable after several model generations.
\item \textbf{Diffusion and workflow accumulation.} Researchers needed time to build trust, prompts, retrieval systems, and local validation practices. Adoption then accelerated nonlinearly.
\item \textbf{Composition change.} AI tools reduced entry costs for authors outside established mathematics networks or encouraged interdisciplinary authors to submit to Mathematics categories.
\item \textbf{Strategic submission and policy response.} Expectations about moderation or endorsement changes may have shifted the timing and number of submissions.
\item \textbf{Other field shocks.} New topics, conferences, evaluation practices, labor market pressure, or other changes could have increased Mathematics submissions through an additional channel.
\end{enumerate}

These mechanisms generate distinguishable predictions. A capability threshold account predicts stronger growth in task types with cheap validators, such as formalization, coding, and enumerative search. A diffusion account predicts repeated use by the same authors and institutions followed by broader dispersion. A composition account predicts changes in first time submitters, geographic origin, prior field, and relations between primary and secondary categories. A policy timing account predicts bunching around announced enforcement dates. Other topical shocks predict concentration in particular subfields without corresponding AI disclosures or task signatures.

The author panel already separates two parts of the composition prediction. High output units with no matched activity in the two preceding windows increased from \AuthorHighNoPriorXXV{} to \AuthorHighNoPriorXXVI{}, while the larger group with at least three prior submissions also expanded. The 2026 tail therefore reflects both entry or recomposition and intensified output among established platform visible units.

\subsection{Three system level scenarios}

\textbf{Productivity diffusion with successful quality adaptation.} Submission volume remains high while reproducibility artifacts, formal checking, and reviewer capacity rise. Correction and retraction rates remain stable after accounting for cohort age. Topic diversity is maintained. In this scenario, AI expands the feasible research frontier.

\textbf{Manuscript inflation and attention scarcity.} Submissions and versions rise faster than verification. Review latency, desk rejection, citation error, and unresolved competing claims increase. Important work receives less expert attention. The paper remains abundant while reliable consensus becomes slower.

\textbf{Individual augmentation with collective contraction.} Adopters publish more and advance faster, but research concentrates around data rich and benchmark friendly questions. Topic entropy, team diversity, and follow on engagement decline. This scenario is compatible with strong private incentives and weak collective welfare.

The scenarios can coexist across fields. Mathematics may experience rapid formalization gains alongside an influx of unverified open problem claims; computational physics may gain reliable code assistance while theoretical interpretation converges. Evaluation should therefore operate at the level of tasks and validation regimes.

\section{Scope, limitations, and next tests}
\label{sec:limitations}

The analysis is designed around archive list information flow. Cross listing makes this the appropriate measure of field exposure and moderation load, while unique global submissions provide the platform level denominator. The new primary category decomposition establishes broad 2026 growth across Mathematics subfields and supplies a sharper account of field attribution for the matched eight month window.

December 2022 supplies a public availability reference for a technology whose capabilities, prices, institutional permissions, and norms evolved continuously. It is not the treatment date in the comparative design. The synthetic control instead uses all observations through December 2025 to identify the 2026 anomaly, and mechanism triangulation interprets its timing.

The synthetic control estimates Mathematics specific divergence relative to an evolving scientific system in which donor archives can share platform and technology changes. The placebo rank of \SCMPlaceboRank{} among \SCMPlaceboN{} and the \SCMSensitivityLow\% to \SCMSensitivityHigh\% range across qualified donor and preperiod designs summarize comparative exceptionalness and design sensitivity. Extending the panel beyond the first eight months of 2026 will sharpen both quantities.

Population studies contribute complementary language and adoption measures. Linking them to stable manuscript identifiers, task disclosures, and later proof verification, reproducibility, correction, citation, and reuse outcomes is the decisive next step for estimating scientific quality and mechanism magnitude. Stratification by access, language, geography, seniority, field culture, and local infrastructure can then identify how gains are distributed across researchers and institutions.

\section{Conclusion}

The 2026 mathematics preprint surge is quantitatively large. Mathematics recorded \MathYTDXXVI{} archive list entries through August, \MathYTDOY\% above 2025. Three complementary comparisons reveal its structure. The raw year over year increase is \MathYTDOY\%; continuation of the archive's own pre-2022 trend gives a \MathExcess\% excess; and the preferred 2026 comparative counterfactual gives a \SCMYTDGap\% Mathematics specific gap. Qualified donor and preperiod designs yield \SCMSensitivityLow\% to \SCMSensitivityHigh\%. Mathematics ranks first among \SCMPlaceboN{} eligible archive placebos, with a finite sample empirical probability of \SCMPlaceboP. Growth across \SubfieldPositiveN{} of \SubfieldTotalN{} primary subfields establishes that the shift is broad. The evidence therefore identifies a large and comparatively unusual 2026 Mathematics production regime within a wider platform expansion.

Population language evidence shows rapid AI diffusion. The pseudonymized author panel independently shows a thicker repeated output tail and stronger entry from low prior activity profiles. Capability studies show expanding performance when search is paired with execution, formal proof, curated retrieval, or expert review. Their joint timing and task structure are consistent with delayed diffusion and capability threshold mechanisms. The comparative design identifies the 2026 anomaly, while this independent triangulation evaluates the AI related explanation. The central transformation is that the cost of producing candidates is falling faster than the cost of establishing trust. Verification, attention, and problem choice therefore become the binding constraints.

Human agency becomes more valuable as machines become more capable. Researchers must own the objective, assumptions, tests, interpretation, and decision to publish. Institutions can amplify this agency through risk based disclosure, verification budgets, linked evidence artifacts, fair moderation, and monitoring that values diversity and reliability alongside throughput. Building the infrastructure for checking and understanding alongside the infrastructure for generation will convert the present productivity shock into a durable expansion of the scientific frontier.

\section*{Data, code, and materials availability}

The public archive panel, official global monthly statistics, pseudonymized author paper records, processed counterfactual series, estimator diagnostics, analysis code, figure generation materials, and editable figure sources are maintained separately from the arXiv compilation source and are available from the corresponding author upon reasonable request. No author names, manuscript titles, abstracts, or arXiv identifiers are retained in the author panel.

\section*{AI assistance disclosure}

Generative AI assisted literature discovery, code implementation, diagnostic design, figure preparation, and language editing under human direction. The human authors conceived the research questions and core ideas, established the analytical framework, selected the data and methods, directed and reviewed all code development, evaluated robustness, interpreted the results, designed the figures and tables, structured the manuscript logic, formulated the conclusions, verified the sources and references, and determined the final scientific expression. The human authors take full responsibility for the research and the submitted manuscript.

\section*{Acknowledgments}

No external funding or conflicts of interest are declared.

\appendix

\section{Archive definitions and measurement semantics}
\label{app:archives}

\begin{table*}[!t]
\centering
\caption{Archive units in the monthly panel}
\begin{tabularx}{\textwidth}{p{3cm}Y}
\toprule
Role & Archives \\
\midrule
Target & \code{math} \\
Physics related & \code{astro-ph}, \code{cond-mat}, \code{gr-qc}, \code{hep-ex}, \code{hep-lat}, \code{hep-ph}, \code{hep-th}, \code{math-ph}, \code{nlin}, \code{nucl-ex}, \code{nucl-th}, \code{physics}, \code{quant-ph} \\
Additional comparison & \code{cs}, \code{econ}, \code{eess}, \code{q-bio}, \code{q-fin}, \code{stat} \\
Descriptive context & Official all arXiv unique monthly submissions \\
\bottomrule
\end{tabularx}
\end{table*}

The query records the number printed by the official archive month list endpoint. Collection checks require exactly 104 months for each of twenty archives, unique archive month pairs, strictly positive counts, and dates from 2018-01 through 2026-08. The official export host supplied the same archive listing content when the main host imposed request throttling.

A limited audit of the first 100 records in the March 2026 Mathematics list found 95 with a Mathematics primary archive, three with a Computer Science primary archive, and two with a Mathematical Physics primary archive. This audit verifies that regular list exposure includes cross listed records. It is not extrapolated as an estimate of the historical cross list share.

Archive list and unique submission counts answer different questions. Let manuscript $p$ appear in set $A_{pt}$ of archive lists in month $t$. The global unique count is $\sum_p1$, whereas the sum of archive lists is $\sum_p|A_{pt}|$. For a specific field $f$, the archive list count $\sum_p\ind\{f\in A_{pt}\}$ measures exposure to that field. The denominator follows the scientific question.

\section{Pseudonymized author output definitions}
\label{app:authors}

The author panel uses unique first submissions carrying at least one Mathematics category. Exact normalized display names are linked within and across matched January--August windows, then stored only as twenty character SHA256 digests. A person using multiple display names appears as multiple pseudonymized units, while identical display names may combine people. Name variant splitting tends to lower measured individual output and homonym merging tends to raise it. The reported quantities therefore describe platform visible author strings under a fixed reproducible rule.

\begin{table*}[!t]
\centering
\caption{Pseudonymized author output in matched January--August windows}
\label{tab:authorannual}
\footnotesize
\setlength{\tabcolsep}{4pt}
\begin{tabular}{r r r r r r r r r}
\toprule
Year & Submissions & Author units & Single author & Five plus & Ten plus & Maximum & Gini & Top one percent \\
\midrule
2023 & 27,703 & 44,432 & 7,402 (26.7\%) & 974 & 70 & 29 & 0.264 & 5.2\% \\
2024 & 30,146 & 47,993 & 7,906 (26.2\%) & 1,150 & 82 & 27 & 0.271 & 5.3\% \\
2025 & \AuthorSubmissionsXXV{} & \AuthorUniqueXXV{} & \AuthorSingleXXV{} (\AuthorSingleShareXXV\%) & \AuthorFivePlusXXV{} & \AuthorTenPlusXXV{} & \AuthorMaximumXXV{} & \AuthorGiniXXV{} & \AuthorTopOneShareXXV\% \\
2026 & \AuthorSubmissionsXXVI{} & \AuthorUniqueXXVI{} & \AuthorSingleXXVI{} (\AuthorSingleShareXXVI\%) & \AuthorFivePlusXXVI{} & \AuthorTenPlusXXVI{} & \AuthorMaximumXXVI{} & \AuthorGiniXXVI{} & \AuthorTopOneShareXXVI\% \\
\bottomrule
\end{tabular}
\end{table*}

Single author counts and affiliation status are distinct. The optional API affiliation field appeared in \AuthorAffiliationObservedXXVI{} of the \AuthorSubmissionsXXVI{} submissions in 2026, and \AuthorExplicitIndependentXXVI{} records explicitly used an independent or unaffiliated label. The latter is a conservative lower bound on explicit self identification. The main analysis therefore uses fully observed single authorship and prior activity rather than imputing institutional status from missing fields.

The Gini coefficient summarizes concentration over each active author unit's submission count. The top one percent statistic is the share of all author paper participations attached to the most active one percent of pseudonymized author units. Prior activity for year $y$ is counted in years $y-2$ and $y-1$ over the same eight months. The categories in Figure~\ref{fig:authoroutput} are no prior activity, one or two prior submissions, and at least three. High output denotes five or more submissions in the current window.

\section{Mathematics subfield decomposition}
\label{app:subfields}

For this decomposition, each unique paper is assigned to its recorded primary \code{math.*} category, which prevents cross category double counting. The matched January--August totals are \SubfieldTotalXXV{} in 2025 and \SubfieldTotalXXVI{} in 2026, an increase of \SubfieldAbsoluteIncrease{} or \SubfieldOverallGrowth\%. This primary category total is narrower than the author panel denominator because the latter includes any submission carrying a Mathematics category. The category codes and primary category metadata come from the official arXiv Atom API \parencite{arxivAPI2026}.

\begin{figure*}[!t]
\centering
\includegraphics[width=0.96\textwidth]{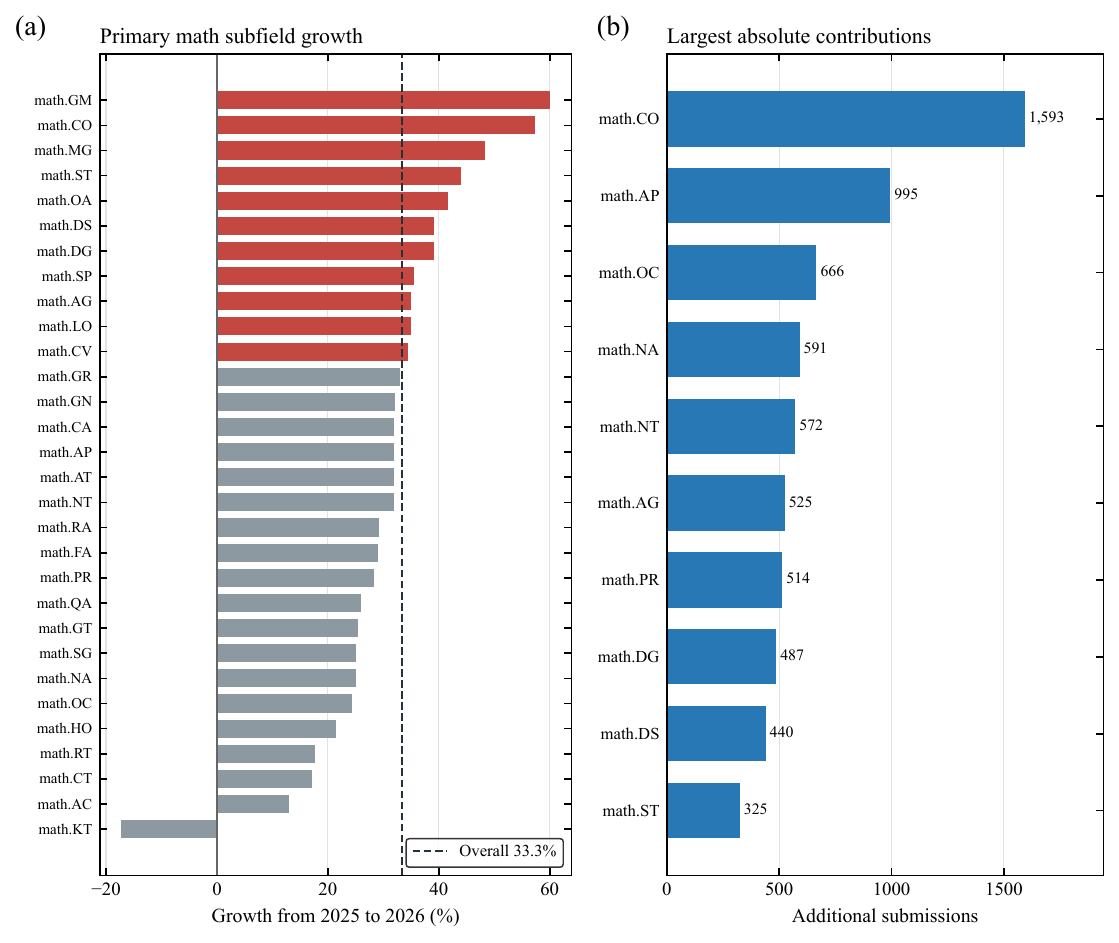}
\caption{Primary Mathematics subfield decomposition for January--August 2025 and 2026 using unique submissions. Panel (a) reports growth for all \SubfieldTotalN{} \code{math.*} primary categories; the dashed line is the overall \SubfieldOverallGrowth\% increase. Panel (b) reports the ten largest absolute contributions to the \SubfieldAbsoluteIncrease{} additional submissions. The direction is broad across categories and the magnitude is heterogeneous.}
\label{fig:subfields}
\end{figure*}

\section{Synthetic control implementation details}
\label{app:scm}

Seasonal adjustment uses calendar month means of log counts calculated on January 2018--December 2025. The same preperiod seasonal factors transform fitted Mathematics logs back to counts. The optimization uses nonnegative weights summing to one and the sequential least squares programming numerical solver, abbreviated SLSQP. Candidate ridge penalties are $\{0,10^{-6},10^{-5},10^{-4},10^{-3},3\times10^{-3},10^{-2},3\times10^{-2},10^{-1}\}$. Six expanding window folds with 12 month validation windows choose the penalty.

The design diagnostics are implemented as follows.
\begin{enumerate}
\item \textbf{Spatial placebos.} Each comparison archive is assigned the January 2026 holdout boundary in turn and ranked by its holdout to preperiod RMSPE ratio after the prefit screen.
\item \textbf{Prior year pseudo holdouts.} The design is reestimated before January--August 2023, 2024, and 2025, with each matched eight month window held out in turn.
\item \textbf{Specification sensitivity.} The preperiod starts in 2018, 2019, or 2020 under the primary donor pool, a pool including \code{math-ph}, a distant field set, and six nonphysics fields.
\item \textbf{Leave one donor out refits.} Every donor receiving at least one percent weight is excluded in turn and the model is reestimated.
\end{enumerate}

\begin{table}[!t]
\centering
\caption{Validation results for the primary synthetic control}
\footnotesize
\begin{tabular}{rrr}
\toprule
Penalty $\lambda$ & Validation MSE & Folds \\
\midrule
0.010 & 0.001537 & 6 \\
0.030 & 0.001580 & 6 \\
0.003 & 0.001607 & 6 \\
0.100 & 0.001682 & 6 \\
0.001 & 0.001685 & 6 \\
0.0001 & 0.001759 & 6 \\
$10^{-5}$ & 0.001781 & 6 \\
$10^{-6}$ & 0.001783 & 6 \\
0 & 0.001784 & 6 \\
\bottomrule
\end{tabular}
\end{table}

The empirical spatial statistic for unit $i$ is
\begin{equation}
R_i=\frac{\sqrt{|T_1|^{-1}\sum_{t\in T_1}(Y_{it}^{*}-\widehat Y_{it}^{*})^2}}
{\sqrt{|T_0|^{-1}\sum_{t\in T_0}(Y_{it}^{*}-\widehat Y_{it}^{*})^2}}.
\end{equation}
After the prefit screen, the finite sample empirical placebo probability is
\begin{equation}
\widehat p=\frac{1}{|\mathcal P|}\sum_{i\in\mathcal P}\ind\{R_i\geq R_M\}
=\frac{1}{15}=0.067.
\end{equation}
This exact finite placebo diagnostic has fifteen eligible units and a corresponding probability resolution of $1/15$.

\section{Weights and sensitivity tables}
\label{app:weights}

\begin{table}[!t]
\centering
\caption{Weights in the primary synthetic Mathematics control}
\footnotesize
\setlength{\tabcolsep}{3pt}
\begin{tabular}{lr@{\qquad}lr}
\toprule
Archive & Weight & Archive & Weight \\
\midrule
\code{astro-ph} & 0.1278 & \code{cond-mat} & 0.1140 \\
\code{hep-th} & 0.1002 & \code{hep-ph} & 0.0918 \\
\code{gr-qc} & 0.0868 & \code{quant-ph} & 0.0811 \\
\code{physics} & 0.0716 & \code{nucl-th} & 0.0712 \\
\code{nucl-ex} & 0.0538 & \code{nlin} & 0.0521 \\
\code{stat} & 0.0487 & \code{q-bio} & 0.0316 \\
\code{q-fin} & 0.0272 & \code{hep-ex} & 0.0216 \\
\code{cs} & 0.0204 & Other donors & $<10^{-5}$ \\
\bottomrule
\end{tabular}
\end{table}

\begin{table}[!t]
\centering
\caption{Donor pool and preperiod sensitivity}
\label{tab:fullsensitivity}
\scriptsize
\setlength{\tabcolsep}{3pt}
\begin{tabularx}{\columnwidth}{Y r r r}
\toprule
Pool and start & Prefit RMSE & Fit ratio & 2026 gap \\
\midrule
Primary 2018 & 0.0339 & 1.00 & 11.9\% \\
Primary 2019 & 0.0309 & 1.00 & 10.6\% \\
Primary 2020 & 0.0293 & 1.00 & 11.6\% \\
Include \code{math-ph} 2018 & 0.0323 & 0.95 & 11.2\% \\
Include \code{math-ph} 2019 & 0.0288 & 0.93 & 9.6\% \\
Include \code{math-ph} 2020 & 0.0262 & 0.89 & 9.9\% \\
Distant 2018 & 0.0334 & 0.99 & 14.7\% \\
Distant 2019 & 0.0330 & 1.07 & 14.5\% \\
Distant 2020 & 0.0307 & 1.05 & 14.9\% \\
Six nonphysics 2018 & 0.1264 & 3.73 & -3.4\%$^{\dagger}$ \\
Six nonphysics 2019 & 0.1027 & 3.32 & -0.7\%$^{\dagger}$ \\
Six nonphysics 2020 & 0.0688 & 2.34 & 11.3\%$^{\dagger}$ \\
\bottomrule
\end{tabularx}

\vspace{0.3em}
\scriptsize $^{\dagger}$Prefit RMSE exceeds twice the primary value with the same start year.
\end{table}

The distant field pool excludes \code{cs}, \code{quant-ph}, \code{stat}, and \code{math-ph}; the six nonphysics pool consists of \code{cs}, \code{econ}, \code{eess}, \code{q-bio}, \code{q-fin}, and \code{stat}. The six nonphysics results demonstrate the identifying role of the prespecified preperiod fit criterion.

\section{Institutional monitoring details}
\label{app:dashboard}

Table~\ref{tab:dashboard} pairs each governance objective with an indicator, cadence, and interpretation rule.

\begin{table}[!t]
\centering
\caption{Proposed monitoring dashboard for AI mediated scientific production}
\label{tab:dashboard}
\footnotesize
\setlength{\tabcolsep}{3pt}
\begin{tabularx}{\columnwidth}{>{\raggedright\arraybackslash}p{1.55cm}Y}
\toprule
Dimension & Indicator, cadence, and interpretation \\
\midrule
Volume & Unique submissions, archive list exposures, and versions per item. Measure monthly by primary and secondary field; separate growth from cross listing and resubmission. \\
AI process & Disclosed use by task and risk tier. Measure monthly or quarterly among submitted and accepted items; track disclosure norms and detection changes. \\
Verification & Formal check coverage, code execution success, independent replication, and time to resolve flags. Measure quarterly by article type and validation regime. \\
Review load & Moderator minutes, invitations per completed report, latency, withdrawals, and rejection reasons. Report monthly medians and tails; interpret latency jointly with review depth. \\
Reliability & Corrections, retractions, citation verification failures, benchmark leakage, and reproducibility failures. Follow annual cohorts to accommodate long lags. \\
Diversity & Topic entropy, semantic distance, new question share, and null result share. Measure annually within field and report sensitivity to the embedding choice. \\
Distribution & Author and institution concentration, geographic access, early career entry, and appeal outcomes. Measure quarterly or annually and pair aggregates with individual level study. \\
Knowledge value & Reuse of code, data, and proofs, together with replication and downstream confirmation. Follow multiyear cohorts and combine citations with reuse. \\
\bottomrule
\end{tabularx}
\end{table}

The indicators should be interpreted jointly. Higher submission volume represents scientific expansion when verification coverage, reliability, diversity, and later reuse rise with it. Cohort tracking connects current process measures to corrections, replication, and downstream confirmation after the relevant publication lags.

Announced policy changes create evaluable interventions. Controlled interrupted designs can compare affected and less affected domains around implementation dates while jointly estimating accepted quality, moderator load, appeals, and representation. Prespecified outcome families and subgroup analyses make productivity and procedural fairness part of the same institutional objective.

\section{Researcher adaptation protocol}
\label{app:protocol}

The task ledger can be implemented through six checkpoints.
\begin{enumerate}
\item \textbf{Question ownership.} State why the problem matters before asking the model how to solve it.
\item \textbf{Assumption inventory.} Record definitions, regimes, symmetries, priors, and exclusion criteria in a human auditable form.
\item \textbf{Adversarial generation.} Request counterexamples, incompatible interpretations, and tests that could refute the proposed result.
\item \textbf{Independent verification.} Apply a different method, implementation, model family, proof checker, limiting case, or human expert.
\item \textbf{Provenance.} Preserve sources, code versions, seeds, environments, material prompts, and rejected alternatives.
\item \textbf{Stopping and disclosure.} Define sufficient evidence and disclose AI involvement at the level required to reproduce or audit the claim.
\end{enumerate}

\section{Evidence map and next tests}
\label{app:audit}

Table~\ref{tab:evidencemap} connects each central claim to its supporting evidence and the next discriminating test.

\begin{table}[!t]
\centering
\caption{Evidence supporting the central claims and their next empirical tests}
\label{tab:evidencemap}
\footnotesize
\setlength{\tabcolsep}{3pt}
\begin{tabularx}{\columnwidth}{>{\raggedright\arraybackslash}p{2.15cm}Y}
\toprule
Claim & Evidence and next test \\
\midrule
Mathematics information flow surged in 2026 & Official monthly archive lists and year over year counts. The next test uses primary category panels and later quality outcomes. \\
Mathematics diverged from weighted donors & Close synthetic prefit, spatial rank, prior year pseudo holdouts, and leave one donor out stability. The next test adds a longer holdout period and enriched archive covariates. \\
The divergence is concentrated in 2026 & Fitted annual gaps remain within two percent through 2025 and reach \SCMYTDGap\% in 2026. The next test extends the holdout and measures monthly author composition. \\
The surge is broad across subfields & Primary category counts increase in \SubfieldPositiveN{} of \SubfieldTotalN{} \code{math.*} categories with heterogeneous magnitudes. The next test links category growth to task and validation regimes. \\
The repeated output tail thickened in 2026 & Pseudonymized official metadata show increases in single author, five plus, and ten plus submission counts and shares. The next test links production profiles to version and validation outcomes. \\
AI contributes to verifiable research steps & Formal proof, executable search, structured calculations, and expert evaluations. The next test applies common research task protocols across systems. \\
AI changes researcher and collective outcomes & Large matched and modeled observational studies. The next test uses prospective adoption designs and institutional variation. \\
Publishing governance is becoming operational & Endorsement expansion, disclosure policy, and monitoring proposals. The next test measures moderator load, audit compliance, appeals, and representation. \\
\bottomrule
\end{tabularx}
\end{table}

The evidence map defines a staged empirical program. Stable manuscript identifiers can connect category histories, submission timing, disclosures, author trajectories, verification artifacts, review outcomes, and later reuse. This linkage directly estimates how much of the 2026 comparative gap is concentrated in documented AI assisted workflows and how those workflows differ in validation intensity.

Institutional variation supplies additional causal leverage. Staggered access to approved tools, training, compute, and endorsement rules can support difference in differences, event studies, and synthetic controls. Mediation analysis can then separate capability, adoption, author composition, and verification pathways, while heterogeneous treatment effects reveal which tasks and researchers gain most from the new production technology.

\FloatBarrier
\printbibliography

\end{document}

%% file: generated_descriptive_metrics.tex
\newcommand{\MathYTDXXVI}{47,127}

\newcommand{\MathYTDOY}{33.5}
\newcommand{\MathExcess}{36.8}
\newcommand{\MathExcessLow}{36.8}
\newcommand{\MathExcessHigh}{62.1}

\newcommand{\PhysicsYTDXXVI}{95,089}

\newcommand{\PhysicsYTDOY}{16.5}
\newcommand{\PhysicsExcess}{18.3}

\newcommand{\GlobalYTDXXVI}{230,322}

\newcommand{\GlobalYTDOY}{26.8}
\newcommand{\GlobalExcess}{36.8}

%% file: generated_causal_metrics.tex
\newcommand{\SCMYTDActual}{47,127}
\newcommand{\SCMYTDSynthetic}{42,113}
\newcommand{\SCMYTDGap}{11.9}

\newcommand{\SCMPreRMSE}{0.034}

\newcommand{\SCMPlaceboRank}{1}
\newcommand{\SCMPlaceboN}{15}
\newcommand{\SCMPlaceboP}{0.067}
\newcommand{\SCMSensitivityLow}{9.6}
\newcommand{\SCMSensitivityHigh}{14.9}

\newcommand{\SCMLeaveOneOutLow}{11.3}
\newcommand{\SCMLeaveOneOutHigh}{13.4}
\newcommand{\SCMGapXXII}{-0.2}
\newcommand{\SCMGapXXIII}{0.7}
\newcommand{\SCMGapXXIV}{1.6}
\newcommand{\SCMGapXXV}{-1.9}
\newcommand{\SCMGapXXVI}{11.9}
\newcommand{\SCMPseudoXXIII}{1.1}
\newcommand{\SCMPseudoXXIV}{1.9}
\newcommand{\SCMPseudoXXV}{-6.6}
\newcommand{\SCMPseudoXXVI}{11.9}

%% file: generated_author_metrics.tex
\newcommand{\AuthorSubmissionsXXV}{31,613}
\newcommand{\AuthorUniqueXXV}{50,302}
\newcommand{\AuthorSingleXXV}{8,450}
\newcommand{\AuthorSingleShareXXV}{26.7}
\newcommand{\AuthorFivePlusXXV}{1,234}
\newcommand{\AuthorFivePlusActiveShareXXV}{2.45}
\newcommand{\AuthorTenPlusXXV}{107}
\newcommand{\AuthorTenPlusActiveShareXXV}{0.21}
\newcommand{\AuthorMaximumXXV}{20}
\newcommand{\AuthorGiniXXV}{0.272}
\newcommand{\AuthorTopOneShareXXV}{5.4}
\newcommand{\AuthorHighNoPriorXXV}{90}
\newcommand{\AuthorHighNoPriorShareXXV}{7.3}
\newcommand{\AuthorHighLowPriorXXV}{318}
\newcommand{\AuthorSubmissionsXXVI}{42,236}
\newcommand{\AuthorUniqueXXVI}{58,930}
\newcommand{\AuthorSingleXXVI}{13,070}
\newcommand{\AuthorSingleShareXXVI}{30.9}
\newcommand{\AuthorFivePlusXXVI}{2,241}
\newcommand{\AuthorFivePlusActiveShareXXVI}{3.80}
\newcommand{\AuthorTenPlusXXVI}{289}
\newcommand{\AuthorTenPlusActiveShareXXVI}{0.49}
\newcommand{\AuthorMaximumXXVI}{44}
\newcommand{\AuthorGiniXXVI}{0.313}
\newcommand{\AuthorTopOneShareXXVI}{6.5}

\newcommand{\AuthorHighNoPriorXXVI}{267}
\newcommand{\AuthorHighNoPriorShareXXVI}{11.9}
\newcommand{\AuthorHighLowPriorXXVI}{696}
\newcommand{\AuthorAffiliationObservedXXVI}{702}
\newcommand{\AuthorExplicitIndependentXXVI}{4}

%% file: generated_subfield_metrics.tex
\newcommand{\SubfieldTotalXXV}{27,102}
\newcommand{\SubfieldTotalXXVI}{36,126}
\newcommand{\SubfieldAbsoluteIncrease}{9,024}
\newcommand{\SubfieldOverallGrowth}{33.3}
\newcommand{\SubfieldPositiveN}{29}
\newcommand{\SubfieldTotalN}{30}
\newcommand{\SubfieldLargestCode}{math.CO}
\newcommand{\SubfieldLargestIncrease}{1,593}
\newcommand{\SubfieldLargestShare}{17.7}

\newcommand{\SubfieldFastLargeGrowth}{57.4}

\newcommand{\SubfieldTopTenShare}{74.3}